\documentstyle[editedvolume,epsfig,numreferences]{crckapb} 

\newcommand{\hep}[1]{{\tt hep-ph/#1}}
\newcommand\epj[3]{{\it Eur. Phys. J. }{\textbf C #1}, p. #2}

\newcommand\jp[3]{{\it J.~Phys. }{\textbf #1}, p. #2}
\newcommand\npb[3]{{\it Nucl. Phys. }{\textbf B #1}, p. #2}
\newcommand\npbps[3]{{\it Nucl. Phys. }
                    {\textbf B} { (Proc. Suppl.)}{ \textbf #1}, p. #2}
\newcommand\plb[3]{{\it Phys. Lett. }{\textbf B #1}, p. #2}                   
\newcommand\prd[3]{{\it Phys. Rev. }{\textbf D #1}, p. #2}            
\newcommand\zpc[3]{{\it Z. Physik }{\textbf C #1}, p. #2}             
\newcommand\sjnp[3]{{\it Sov. J. Nucl. Phys. }{\textbf #1}, p. #2}    
\newcommand\jetp[3]{{\it Sov. Phys. JETP }{\textbf #1}, p. #2}        
 
\newcommand{\as}{\alpha_{\mathrm{s}}}

\newcommand{\cut}{_{\rm cut}}
\newcommand{\dJ}[1]{\frac{\dif\sigma}{\dif J}{}\raisebox{1.5ex}{$#1$}}
\newcommand{\dif}{{\rm d}}
\newcommand{\dk}{\dif\kk}
\newcommand{\dug}{\,\raisebox{0.37pt}{:}\hspace{-3.2pt}=}
\newcommand{\e}{\varepsilon}
\newcommand{\esp}[1]{{\rm e}^{#1}}
\newcommand{\fz}{f^{(0)}}
\newcommand{\hz}{h^{(0)}}
\newcommand{\id}{\mbf{1}}

\newcommand{\kd}{\kk_2}
\newcommand{\kk}{\mbf{k}}
\newcommand{\ku}{\kk_1}

\newcommand{\mbf}[1]{\mbox{\boldmath $#1$}}

\newcommand{\pa}{{\mathsf a}}
\newcommand{\pb}{{\mathsf b}}
\newcommand{\pg}{{\mathsf g}}

\newcommand{\pq}{{\mathsf q}}
\newcommand{\product}{*}
\newcommand{\qq}{\mbf{q}}

\newcommand{\ui}{{\mathrm i}}

\newcommand{\lab}[1]{\label{#1}}
\newcommand{\labe}[1]{\label{#1}}
\begin{opening}
\title{
\centerline{Jet Vertex in the Next-to-Leading log(s)}
\centerline{approximation}
}

\author{G. P. Vacca}
\institute{Dipartimento di Fisica, Universit\`a di Bologna and \\
INFN, Sez. di Bologna, via Irnerio 46, 40126 Bologna, Italy\\}

\end{opening}

\runningtitle{Jet Vertex in the Next-to-Leading log(s) approximation}

\begin{document}

\begin{abstract}
The next-to-leading corrections to the jet vertex which is relevant for
the Mueller-Navelet jets production in hadronic collisions and for the
forward jet cross section in lepton-hadron collisions are presented in the
context of a $k_t$ factorizazion formula which resums the leading and
next-to-leading logarithms of the energy.
\end{abstract}

\section{Introduction}
Recently, in the study of QCD in the Regge limit,
a novel element has been defined and computed at the
next-to-leading order (NLO). It is the jet vertex \cite{BaCoVa},
which represents one of the building
blocks in the production of Mueller-Navelet jets at hadron
hadron  colliders and of  forward jets~\cite{MuNa87}
in deep inelastic electron proton scattering. 
Such processes should provide a kinematical environment
for which the BFKL Pomeron~\cite{BFKL76} QCD analysis could
apply, provided that the transverse energy of the jet fixes a
perturbative scale and the large energy yields a large rapidity interval.

In a strong
Regge regime important contributions, or even dominant, come, in the
perturbative language, from diagrams beyond NLO and NNLO at fixed order in 
$\alpha_s$. This is the main reason for considering a resummation of the
leading and next-to-leading logarithmic contributions as
computed in the BFKL Pomeron framework. The lackness of
unitarity, if the related corrections are not taken into account,
forces one to consider an upper bound on the energy to suppress them.
It is already known that the leading logarithmic (LL) analysis is not
accurate enough~\cite{nonasympt}, being the
kinematics selected by experimental cuts far from any asymptotic regime.
Moreover at this level of accuracy there is a maximal dependence in the
different scales involved (renormalization, collinear factorization
and energy scales). For the Mueller-Navelet jet production process the only
element still not known at the NLO level was the ``impact factor'', which
describes the hadron emitting one inclusive jet when interacting with the
reggeized gluon which belongs to the BFKL ladder, accurate up to
NLL~\cite{BFKLNLO}. The jet vertex, now computed,
is the building block of this interaction.
For the so called forward jet production in DIS the extra ingredient
necessary is the photon impact factor, whose calculation is currently
in progress~\cite{BaGiQi00,FaMa99}.  
Let us also remind that NLL BFKL approach has recently
gained more theoretical solidity since the bootstrap condition in its strong
form, which is the one necessary for the self-consistency of the
assumption of Reggeized form of the production
amplitudes, has been stated and
formally proved~\cite{bootstrapnlo}. This relation is a very remarkable
property of QCD in the high energy limit.

\begin{figure}[hb!]
\centering
\resizebox{0.3\textwidth}{!}{\includegraphics{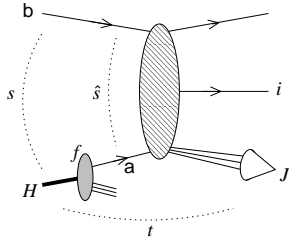}}
\caption{\labe{f:jet} High energy process with jet production.
$H$ is the incoming hadron providing a parton $\pa$ (gluon $\pg$/quark $\pq$)
with distribution density $f_\pa$ which scatters with the parton $\pb$
with production of a jet $J$ in the forward direction (w.r.t $H$) and
$i$ is the generic label for outgoing particles.}
\end{figure}

A theoretical challenge, interesting by itself and
appearing in the calculation, is related to the special kinematics.
The processes to be analyzed is illustrated in Fig.\ref{f:jet}:
the lower parton
emitted from the hadron $H$ scatters with the upper parton
$\pq$ and produces the jet $J$.
Because of the large transverse momentum of the jet 
the parton is hard and the collinear factorization allows for a partonic
scale dependence described by the DGLAP evolution equations \cite{DGLAP}.
Above the jet, on the other hand, the kinematics chosen requires a
large rapidity gap between  the jet and the outgoing parton $\pq$:
such a situation is described by BFKL dynamics.
Therefore the jet vertex lies at the interface between
DGLAP and BFKL dynamics, a situation which appears for the first time in
a non trivial way. As an essential result of our analysis we find 
that it is possible to separate, inside the jet vertex, the collinear infrared 
divergences that go into the parton evolution of the incoming gluon/quark from 
the high energy gluon radiation inside the rapidity gap which belongs to 
the first rung of the LO BFKL ladder.

\section{Jet Vertex: definition and outline of the derivation}
Let us consider, for the process illustrated in Fig. \ref{f:jet},
the kinematic variables
\begin{eqnarray}
p_H &=& \left(\sqrt{s/2},0,\mbf{0}\right)\quad , \quad p_\pa = x\,p_H
\quad , \quad
p_\pb = \left(0,\sqrt{s/2},\mbf{0}\right) \nonumber \\
p_i &=& E_i\left(\esp{y_i}/\sqrt2,\esp{-y_i}/\sqrt2,\mbf{\phi}_i\right)
\quad , \quad s\dug(p_H+p_\pb)^2 \; .
\end{eqnarray}
We study the partonic subprocess $\pa+\pb\to X+{\it
jet}$ in the high energy limit
\begin{equation}\lab{HElimit}
\Lambda_{\rm QCD}^2 \ll E_J^2\sim -t\; {\rm (fixed)} \ll s\to\infty.
\end{equation}
Starting from the parton model, we assume the physical cross section to be
given by the corresponding partonic cross section $\dif\hat\sigma$
(computable in perturbation theory)
convoluted with the parton distribution densities (PDF) $f_\pa$ of the
partons $\pa$ inside the hadron $H$.
A jet distribution $S_J$, with the usual infrared safe
behaviour, selects the final states contributing to the one
jet inclusive cross section that we are considering.
We choose, as jet variables, rapidity, transverse energy and azimuthal
angle and the one jet inclusive cross section initiated by quarks and
gluons in hadron $H$ is therefore written as
\begin{equation}\lab{pSf}
\dJ{} \dug \frac{\dif\sigma_{\pb H}}{\dif y_J \dif E_J \dif \phi_J}
= \sum_{\pa=\pq,\pg} \int\dif x\;
\dif\hat\sigma_{\pb\pa}(x)\,S_J(x)\fz_\pa(x)\;.
\end{equation}

At the {\it lowest order} the jet cross section,
dominated by a $t$-channel gluon exchange,
can be written as~\cite{BaCoVa}
\begin{equation}\lab{LOFF}
 \dJ{^{(0)}} = \sum_{\pa=\pq,\pg}
\int\dif x \int \dif\kk\;\hz_\pb(\kk)V^{(0)}_\pa(\kk,x)\fz_\pa(x)
\end{equation}
where $V^{(0)}_\pa(\kk,x)=\hz_\pa(\kk)
S_J^{(2)}(\kk,x)$ is the jet vertex induced
by parton $\pa$, $\hz_\pa(\kk)$ is the partonic impact
factor (generally expressed in $D=4+2\e$ dimensions for
regularization pourposes at the NLO)
and $\fz_\pa(x)$ is the parton distribution density (PDF).
The jet distribution is in this case trivial
\begin{equation}
S_J^{(2)}(\kk,x)\dug S_J^{(2)}(p_1,p_2;p_\pg,p_\pq)= \delta\left(1-x_J/x
\right)E_J^{1+2\epsilon} \delta(\kk-\kk_J)
\end{equation}
with
$ x_J \dug E_J\esp{y_J}/\sqrt s$.

In the {\it NLO approximation} virtual and real corrections enter in the
calculation of the partonic cross section
$\dif\hat\sigma_{\pb\pa}$. The three partons, produced in the real
contributions, in the upper, and lower rapidity region are denoted by
$2$ and $1$ respectively, while the third, which can be emitted
everywhere in rapidity, by
$3$. Moreover we shall call $k=p_\pb-p_2$ and $k'=p_1-p_\pa$ and
$q=k-k'$.
A Sudakov parametrization is chosen 
\begin{eqnarray}
&& k = -\bar{w} p_\pg + w p_\pq + k_\perp\; \quad , \quad
 k_\perp = (0,0,\kk)\\ \label{k2}
&& k' = -z p_\pg + \bar{z} p_\pq + k'_\perp\; \quad , \quad
 k'_\perp = (0,0,\kk')\;
\end{eqnarray}
with the typical inequalities $\bar{w} \sim \bar{z} \ll w \sim z \ll
1$ valid in the multi Regge kinematics (MRK).
Dimensional regularization is used, as usual, to trace
the infrared and ultraviolet divergences.
On taking into account the infrared (IR) properties of the jet distribution
$S_J^{(3)}$,  the following structure is matched exactly up to NLO
(i.e. $\alpha_s^3$)~\cite{BaCoVa}
\begin{equation}
\dJ{}= \sum_{\pa=\pq,\pg} \int\dif x \int \dif\kk\,\dif\kk'\;
 h_\pb(\kk) G(xs,\kk,\kk') V_\pa(\kk',x) f_\pa(x) \; ,
\label{jetcrosssection}
\end{equation}
where
\begin{eqnarray}
&&h = \hz + \as h^{(1)}+\cdots\; , \nonumber \\
&&V = V^{(0)} + \as V^{(1)}+\cdots\; , \nonumber \\
&&f = \fz + \as f^{(1)}+\cdots \; , \nonumber \\
&&G(xs,\kk,\kk') \dug \delta(\kk-\kk')+
\as K^{(0)}(\kk,\kk')\log\frac{xs}{s_0}+\cdots
\end{eqnarray}
The partonic impact factor correction in forward direction $h^{(1)}$
is well known~\cite{CiCo98}, the PDF's $f_\pa$ are the standard ones
satisfying DGLAP evolution equations
\begin{equation}
 \as f^{(1)}_\pa(x,\mu_F^2) \dug
\frac\as{2\pi}\,\frac1\e\left(\frac{\mu_F^2}{\mu^2}\right)^\e 
 \sum_\pb P_{\pa\pb}\otimes \fz_\pb\,
\end{equation}
with $\mu_F$ the collinear factorization scale,
and the BFKL Green's function
$G$ is defined by the LO BFKL kernel $K^{(0)}$. The new element
is the correction to the jet vertex $V^{(1)}$ whose
expressions can be found in~\cite{BaCoVa}. 

Let us note that the previous form in (\ref{jetcrosssection}) is
clearly suggested by the structure of the leading logarithmic part, 
but at this level it is not a trivial ansatz. Only after a proper treatment of
all the IR singular terms one will be able to extract and define the
jet verteces initiated by quarks and gluons in the hadron.

The NLO virtual corrections in the high energy limit are relatively
simple and they can be summarized in the following expression:
\begin{eqnarray}\nonumber
 \dJ{^{(\rm virt)}} =&& \!\!\!\!\! \as \sum_a \int\dif x\int\dk\;\hz_\pq(\kk)
 \left[2\omega^{(1)}(\kk)\log\frac{xs}{\kk^2}
 +\widetilde\Pi_\pb(\kk)+\widetilde\Pi_\pa(\kk)\right]\\ \lab{dJvirt}
 && \!\!\!\!\! \times V^{(0)}_\pa(\kk,x) \fz_\pa(x)\;,
\end{eqnarray}
where $\omega^{(1)}(\kk)$ is the 1-loop reggeized gluon trajectory and
the $\widetilde\Pi_i(\kk)$ are related to the part constant in energy,
after having performed the ultraviolet (UV) $\e$-pole subtraction
occuring in the renormalization of the coupling.

The NLO real corrections, even in the high energy limit, are much more
involved since the interplay of different IR soft and collinear
singularities follows a more complicated pattern.

One starts from the partonic differential cross sections
$\dif\hat\sigma_{\pb\pa}(x)$, appearing in (\ref{pSf}), which have been
also computed in the high energy regime~\cite{CiCo98}, since in such a case
all terms suppressed by powers of $s$ may be neglected.
The form of the partonic differential cross section turns out to be
quite simple when restricted to one of the two halves of the phase
space, which are obtained introducing the rapidity $y'=y+\frac12\log\frac1x$
``measured'' in the partonic center of mass frame, and
splitting the phase space into two semi\-spaces defined by $y'_1, y'_3 > 0$
({\em lower half}) and $y'_3<0<y'_1$ ({\em upper half}).

In the quark initiated case one obtains for both regions a reasonable
simple expression, while for the gluon initiated part it is convenient
to consider separately the $\pq\bar{\pq}$ and $\pg\pg$ final state cases.

The ``upper half region'', with $y'_3<0$, leads to the rederivation of
the partonic $\pb$-impact factor, already well known at NLL.
The ``lower half region'', wherein $y_3'>0$, which corresponds to
$z>z\cut\dug\frac{E_3}{\sqrt{xs}}$ is precisely the one from which we
are able to extract the jet verteces.

A very important ingredient, as previosly mentioned, is given by the
jet definition $S_J^{(n)}$. As usual it selects from a generic
$n$-particle final state the configurations contributing to our one
jet inclusive observable and it must be IR safe.
The last requirement simply corresponds to the fact that emission of a
soft particle cannot be distinguished from the analogous state without
soft emission. Furthermore, collinear emissions of partons cannot be
di\-stin\-gui\-shed from the corresponding state where the
collinear partons are replaced by a single parton carrying the sum of
their quantum numbers.

The extraction of all the IR singularities
from the real contributions and the matching with the virtual ones
is done on employing the standard subtraction method.
This consists in approximating the amplitudes in
any singular region in order to extract the exact analytic singular
behavior, and in leaving the remaining finite part in a form suitable
for numerical integration, since normally a full analytical
computation is tecnically impossible.
The analytic singular terms of both virtual and real contributions
are therefore combined. Moreover, with respect to standard NLO jet
analysis, one has also to deal with the leading logarithmic terms
which are entangled to the other IR singular terms. 
For the process considered it can be
shown~\cite{BaCoVa} that finally one is left with the collinear
singularities associated to the impact factor of the parton $\pb$, to the
definition of the distribution function for the parton $\pa$ in $H$
and to the BFKL kernel term. 
The finite remaning parts correspond to the impact factors and jet verteces.

We recall that the jet distribution functions become essential in
disentangling the collinear singularities, the soft singularities,
and the leading $\log s$ pieces when the particle $3$ is a gluon.
The same basic mechanism may be used for both the quark and the gluon
initiated cases.
\begin{itemize}
\item When the outgoing parton $1$ is in the collinear region of the incoming parton $\pa$,
  i.e., $y_1\to\infty$, it cannot enter the jet; only gluon 3 can thus be the jet, $y_3$ is
  fixed and no logarithm of the energy can arise due to the lack of evolution in the gluon
  rapidity. No other singular configuration is found when $J=\{3\}$.
\item In the composite jet configuration, i.e., $J=\{1,3\}$, the gluon rapidity is bounded
  within a small range of values, and also in this case no $\log s$ can arise. There could
  be a singularity for vanishing gluon $3$ momentum: even if the $1\parallel3$ collinear
  singularity is absent, we have seen that, at very low $z$, a soft singular integrand
  arises. However, the divergence is prevented by the jet cone boundary, which causes a
  shrinkage of the domain of integration $\sim z^2$ for $z\to0$ and thus compensates the
  growth of the integrand.
\item The jet configuration $J=\{1\}$ corresponds to the situation
  wherein gluon $3$ spans the whole phase space, apart,
  of course, from the jet region itself. The LL term arises from gluon
  con\-fi\-gu\-ra\-tions in the central region. Therefore, it is crucial to understand to
  what extent the differential cross section provides a leading contribution. It turns out
  that the coherence of QCD radiation suppresses the emission probability for gluon $3$
  rapidity $y_3$ being larger than the rapidity $y_1$ of the parton $1$, namely an angular
  ordering prescription holds. This will provide the final form of the leading term, i.e.,
the appropriate scale of the energy and, as a consequence, a finite
and definite expression for the one-loop jet vertex correction.
\end{itemize}

It is therefore clear that the energy scale $s_0$
associated to the BFKL rapidity evolution plays a crucial role.
The calculations show a
natural choice, due to angular ordered preferred gluon emission and
the presence of the jet defining distribution, which
is also crucial to obtain the full collinear singularities which
factorize into the PDF's.
In particular, on considering the term giving the real part of the LL
contribution, one can see that outside the angular ordered region
\begin{equation}\label{angordpos}
 \frac{E_3}{z} > \frac {E_1}{1-z} \quad\iff\quad \theta_3 > \theta_1 
 \quad\iff\quad y_3 < y_1\;,
\end{equation}
there is a small contribution to the cross section.
This ordering sets the natural energy scale
which, for example, leads to the expression of the jet vertex for
the case $s_0(\kk,\kk') \dug (|\kk'|+|\qq|)(|\kk|+|\qq|)$.
A mild modification of such a scale can be
performed without introducing extra singularities, contrary to what
happens for a generic choice.
In any case the use of a different scale requires the
introduction of modifing terms. For the useful
symmetric Regge type energy scale $s_R=|\kk||\kk'|$, one has
\begin{equation}
 G(xs,\kk,\kk')=(\id+\as H_L)\left[\id+\as K^{(0)}
\log\frac{xs}{|\kk||\kk'|}\right]
(\id+\as H_R),
\end{equation}
where
$H_L(\kk,\kk')=-K^{(0)}(\kk,\kk')\log\left(|\kk|+|\qq|)/|\kk|\right)=
H_R(\kk',\kk)$.

We do not present the final vertex expressions here, they can be found
in \cite{BaCoVa}.

\section{Remarks and final formulas}
In order to extend the results of the two loop calculations, which
have allowed to obtain the full NLL partonic cross sections,
to our inclusive jet production case, it is necessary to check the
relation between the partonic impact factors and the jet verteces.
More precisely one may define the one loop correction to the
``jet impact factor'' as
$[V\product f]^{(1)}=V^{(1)}\product\fz+V^{(0)}\product f^{(1)}$.
This latter object and the corrections to the partonic impact factor
$ h^{(1)}$ are different in one important aspects, i.e. their
collinear singularities.
Essentially the IR singularity associated to the LL real term is absent
(subtracted) in the partonic impact factors, while this does not happen
in the ``jet impact factors'' due to the constraining presence of
the jet. In the ``jet impact factors'' are therefore hidden the full
collinear singularities which should be factorized to obtain the right
behavior of the PDF's satisfying the DGLAP evolution equations.

Apart from this important difference there is a full correspondence,
as shown in \cite{BaCoVa}, between these two objects.
In particular the same angular ordering mechanism, which sets
a reference for the energy scale, is present. Thus
any tuning of the BFKL kernel energy scale is done precisely in the
same way for both impact factors and jet vertices and, thanks to the
Regge factorization property valid up to NLL in QCD,
one is allowed to use the known form of the NLL BFKL Green's function
setting the scale to $s_R$, as previously discussed.

From a two loop analysis, up to $\alpha_s^4$ order, one expects to
find the contributions
\begin{eqnarray}\nonumber
 \frac1{\as^4}\dJ{^{(2)}} \!\!\! &=& \!\! \frac12\log^2\frac{xs}{s_R}\;\hz
K^{(0)} K^{(0)} V^{(0)} \fz +
\log\frac{xs}{s_R}\Big[h^{(1)} K^{(0)} V^{(0)} \fz \\ \nonumber
&&+\hz\big(H_L K^{(0)}+ K^{(0)} H_R+K^{(1)}\big) V^{(0)} \fz
+\hz K^{(0)} V^{(1)} \fz\\ \label{2loop}
&&+\hz K^{(0)} V^{(0)} f^{(1)}\Big]\;,
\end{eqnarray}
where the first term on the RHS is LL while the terms collected
in the square brackets are NLL.

Consequently, to obtain the jet cross section with accuracy up to
NLL terms, one has to consider the NLL BFKL kernel
\begin{equation}
K = \as K^{(0)}+\as^2 K^{(1)}
\end{equation}
which has been computed with the Regge type scale $s_R=|\kk_1||\kk'_2|$ and
leads to a Green's function, which resums all terms up to NLL
in (\ref{jetcrosssection}),
\begin{equation}
G(xs,\ku,\kd) =
\int\frac{\dif\omega}{2\pi\ui}\left(\frac{xs}{s_R}\right)^\omega
\langle\ku|(\id+\as H_L) [\omega - K]^{-1}(\id+\as H_R)|\kd\rangle\,.
\label{fullG}
\end{equation}
The formula for the Mueller-Navelet jets~\cite{BaCoVa} can be easily
derived symmetrizing the formula (\ref{jetcrosssection}) for the two
jet case.
More precisely one obtains
\begin{eqnarray} \lab{finaljet}
\frac{\dif^2 \sigma}{\dif J_1\dif J_2} &=& \sum_{\pa,\pb} \int \dif x_1 \dif x_2
  \int \dif \kk_1 \dif \kk_2\; f_{\pa}(x_1) V_{\pa}(\kk_1,x_1) 
\nonumber \\ 
&&G(x_1 x_2 s,\kk_1,\kk_2) V_{\pb}(\kk_2,x_2) f_{\pb}(x_2) 
\end{eqnarray} 
where $\pa,\pb=\pq,\pg$, and the subscripts $1$ and $2$ refer to jet
$1$ and $2$ in hadron 
 $1$ and $2$, resp. Once the definition of the
jets is given, all the elements are known and a definitive computation
may be carried on.

Regarding the Green's function, two more remarks should be made.
Since the high energy asymptotic regime is far, a few iterations of
the kernel instead of a full resummation, as given in eq. (\ref{fullG}),
may be of some interest in a phenomenological application.
The known large corrections of the BFKL kernel are related to the
presence of large logs. This problem has been solved using optimal
renormalization scale methods \cite{Brodsky} or by considering an
improved version of the kernel, equivalent at NLL level, but including
correcting higher order terms, following a collinear prescription
\cite{CiaCo1999}.

\section{Conclusions}
A brief review of the definitions and calculations at the NLO accuracy
of the jet vertex relevant for the Mueller-Navelet jets at
hadron-hadron colliders and for forward jets in deep inelastic
electron-proton scattering  has been given.
This object, defined for high energies, due to specific kinematics, lies at
the interface between DGLAP and BFKL dynamics.
Being known all the ingredients, it is now possibile to perform a
numerical analysis of the production of Mueller-Navelet jets at NLL.
For the forward jet processes further theoretical studies on the NLL photon impact
factors are still required.  

\section*{Acknowledgments}
The results discussed have been obtained in collaboration with J. Bartels
and D. Colferai~\cite{BaCoVa}.

%
\end{document}